\documentclass[preprint,letterpaper,sort&compress,12pt]{elsarticle}

\usepackage[latin9]{inputenc}
\usepackage{amsmath}
\usepackage{amssymb}
\usepackage{graphicx}
\usepackage{verbatim}
\usepackage{bm}
\usepackage{mathbbol}
\usepackage{color}
\setcitestyle{compress}

\usepackage{mathtools}

\usepackage{graphicx,epsfig,amsfonts,amssymb}
\usepackage{bm}
\usepackage{times}
\usepackage{lipsum}
\usepackage{verbatim}

\newcommand{\da}{\downarrow}
\newcommand{\ua}{\uparrow}
\newcommand{\dis}{\displaystyle}

\def\proj#1{|#1\rangle \langle #1|}
\def\out#1#2{|#1\rangle \langle #2|}
\def\g{\gamma}
\def\ve{\varepsilon}

\usepackage{hyperref}
\hypersetup{
    colorlinks,
    citecolor=blue,
    filecolor=blue,
    linkcolor=blue,
    urlcolor=blue
}

\newcommand{\beg}{\begin{equation}}
\newcommand{\en}{\end{equation}}

 \newcommand{\lam}{\lambda}

\newcommand{\eref}[1]{Eq.~(\ref{#1})}
\newcommand{\re}[1]{(\ref{#1})}

\newcommand{\esref}[1]{Eqs.~(\ref{#1})}

\newcommand{\eps}{\varepsilon}

\newcommand{\be}{\begin{eqnarray}}
\newcommand{\ee}{\end{eqnarray}}

\newcommand{\bs}{\begin{equation}\begin{split}}
\newcommand{\es}{\end{split}\end{equation}}

\date{\today}

\begin{document}
\title{Integrable   time-dependent Hamiltonians, solvable Landau-Zener models and Gaudin magnets}

\author{Emil A. Yuzbashyan\footnote{Email: eyuzbash@physics.rutgers.edu}}

\address{Department of Physics and Astronomy, Rutgers University, Piscataway, NJ 08854, USA}

\begin{abstract} We solve the non-stationary Schr\"odinger equation for several  time-dependent  Hamiltonians, such as the BCS Hamiltonian with an interaction strength inversely proportional to time,  periodically driven BCS and linearly driven inhomogeneous Dicke models as well as various multi-level Landau-Zener tunneling models. The latter are  Demkov-Osherov, bow-tie, and generalized bow-tie models. We show that  these Landau-Zener problems  and their certain interacting many-body generalizations map to Gaudin magnets in a magnetic field. Moreover, we demonstrate that the time-dependent Schr\"odinger equation for   the above models has a similar structure and is  integrable with a similar technique as Knizhnikov-Zamolodchikov equations. We also  discuss applications of our results to the problem of molecular production in an atomic Fermi gas swept through a Feshbach resonance and to the evaluation of the Landau-Zener transition probabilities.

\end{abstract}

\maketitle

\section{Introduction}

Since the discovery of quantum mechanics, from the Bohr atom and the harmonic oscillator, to the present day,  integrable models have played a key role in our understanding of physics at the quantum level.  The field has acquired a new prominence in nonequilibrium many-body physics with direct observation of signatures of integrable dynamics in cold atom and solid state experiments \cite{kinoshita2006,Gring:2012,shimano2013,Langen:2015} and the realization that quantum integrable systems display properties characteristic of the many-body localized phase of matter \cite{serbyn,huse,ros,imbrie,vasseur}. Nevertheless,  despite great interest in nonequilibrium phenomena,  the vast majority of studies of  integrable many-body interacting systems out of equilibrium    leave out  models of the Landau-Zener type  where one of the couplings or external fields in the Hamiltonian  varies in time in a continuous fashion.   

We consider several examples  of time-dependent models in this paper. The list includes Bardeen-Cooper-Schiffer (BCS) Hamiltonian with a coupling inversely proportional to time as well as periodically driven BCS models, the problem of molecular production in an atomic Fermi gas swept through a Feshbach resonance (driven inhomogeneous Dicke model), and various   models of multi-level Landau-Zener tunneling. Among the latter are Demkov-Osherov \cite{do,be}, bow-tie \cite{bow-tie}, and generalized bow-tie Hamiltonians
\cite{gbow-tie1,gbow-tie} as well as their many-body extensions. Some of these models have been analyzed by other means before,  others, e.g., the time-dependent BCS model, are new. We will see that    they are all closely related, construct  exact general solutions of  their non-stationary Schr\"odinger equations, and explain how they fit into the existing theory of quantum integrability.   This work builds in part on Ref.~\citealp{yuzbashyan-LZ}, where nontrivial  commuting partners
for  solvable Landau-Zener type Hamiltonians have been derived; some of our results have been  briefly announced in Ref.~\citealp{sinitsyn2017}.

  The BCS and Dicke models are known to be related to  Gaudin magnets \cite{gaudin,integ,ortiz2005}, i.e. $N$ commuting spin  Hamiltonians of the form
\beg
\hat H_j^G=2B\hat s_j^z-\sum_{k\ne j}\frac{\hat {\bf s}_j\cdot \hat {\bf s}_k}{\eps_j-\eps_k},\quad [\hat H_i^G, \hat H_j^G]=0,\quad  i, j = 1, \ldots, N,
\label{gaudin}
\en
where $\hat {\bf s}_k$ are quantum spins of arbitrary magnitude $s_k$, and $B$ and $\eps_k$ are arbitrary real parameters. Amazingly, we find that \textit{all} other models listed above also map to Gaudin magnets.
Associated with Gaudin Hamiltonians are differential equations
of the evolution type known as Knizhnik-Zamolodchikov (KZ) equations \cite{KZ},
\beg
i\nu\frac{\partial\Psi_\mathrm{KZ}}{\partial\eps_j}=\hat H_j^G\Psi_\mathrm{KZ}=\biggl(2B\hat s_j^z-\sum_{k\ne j}\frac{\hat {\bf s}_j\cdot \hat {\bf s}_k}{\eps_j-\eps_k}\biggl)\Psi_\mathrm{KZ},
\label{kz}
\en
where $\nu$ is a real parameter and by construction $\Psi_\mathrm{KZ}=\Psi_\mathrm{KZ}(\bm \eps, B)$ is a function of  $(\eps_1,\dots,\eps_N)\equiv \bm\eps$ and $B$. It turns out that  the non-stationary Schr\"odinger equation for each of our examples  has the same structure as  KZ equations~\re{kz} taken in a certain limit or a particular case. This allows us to transfer known results and machinery developed for the KZ equations to these and other similar time-dependent models. 

One such result is an integral representation of the general solution of KZ equations \cite{babujian2}.  In the original KZ equations $B=0$, i.e. the first term inside the brackets in \eref{kz} is absent.  However, it is not difficult to generalize the solution to the $B\ne 0$ case \cite{sierra,amico1,sedrakyan,gritsev}. We add to this an important observation (see Sect.~\ref{bcs-sect}) that the evolution of $\Psi_\mathrm{KZ}$ with $B$ is governed by the BCS Hamiltonian
\beg
i\nu\frac{\partial\Psi_\mathrm{KZ}}{\partial B}=\hat H_\mathrm{BCS}\Psi_\mathrm{KZ},
\label{kzB}
\en
where $\hat H_\mathrm{BCS}$ is the BCS (Richardson) Hamiltonian expressed in terms of Anderson pseudospins  \cite{anderpseudo},
\beg
\hat H_\mathrm{BCS}=\sum_{j=1}^N2\eps_j \hat s_j^z - \frac{1}{2B}\sum_{j,k} \hat s_j^+ \hat s_k^-,
\label{bcs}
\en
and $(2B)^{-1}$ plays the role of the BCS coupling constant. Taking $B$ to be a function of time, $B=B(t)$, turns \eref{kzB} into the non-stationary Schr\"odinger equation for the Hamiltonian
\beg
\hat H_\mathrm{BCS}(t)=\nu^{-1}\dot B\sum_{j=1}^N2\eps_j \hat s_j^z - (2\nu B)^{-1}\dot B\sum_{j,k} \hat s_j^+ \hat s_k^-,
\label{bcs1}
\en
 whose solution is $\Psi_\mathrm{KZ}(\bm\eps, B(t))$. In particular, the choice $B(t)=\nu t$ yields a BCS Hamiltonian with the coupling constant inversely proportional to time,
 \beg
\hat H_\mathrm{BCS}(t)= \sum_{j=1}^N 2\eps_j \hat s_j^z - \frac{1}{2\nu t}\sum_{j,k} \hat s_j^+ \hat s_k^-,
\label{bcs2}
\en
while a periodic $B(t)$ leads to an integrable Floquet BCS superconductor.

In the limit where the length of one of the spins, say ${\bf s}_N$, goes to infinity, so that it is replaced with a harmonic oscillator, $\hat H_N^G$ becomes the inhomogeneous
Dicke (Tavis-Cummings) model \cite{gaudin}
\be
\hat{H}_\mathrm{D}= \sum_{j=1}^{N-1} \xi_j \hat{s}_j^z -\omega \hat n_b + g \!\sum_{j=1}^{N-1}(\hat{b}^{\dagger} \hat{s}^-_j + \hat{b} \hat{s}^{\dagger}_j ),
\quad \hat n_b=\hat{b}^{\dagger} \hat{b},
\label{TC}
\ee
where $\hat b^\dagger$ and $\hat b$ are bosonic creation and annihilation operators. The rest of Gaudin magnets yield its commuting partners   (see Sect.~\ref{D-sect} for details). The corresponding ($j=N$) KZ equation \re{kz}  is replaced with
\beg
i\nu\frac{\partial\Psi_D}{\partial\omega}=\hat H_D\Psi_\mathrm{D}.
\label{dickekz}
\en
For $\omega=\nu t$ this is the non-stationary Schr\"odinger equation for the time-dependent  Hamiltonian
\be
\hat{H}_\mathrm{D}(t)= \sum_{j=1}^{N-1} \xi_j \hat{s}_j^z -\nu t \hat n_b + g \!\sum_{j=1}^{N-1}(\hat{b}^{\dagger} \hat{s}^-_j + \hat{b} \hat{s}^{\dagger}_j ).
\label{TC1}
\ee
In this way we derive in Sect.~\ref{D-sect}  a full set of solutions of the non-stationary Schr\"odinger equation for $\hat{H}_\mathrm{D}(t)$ retracing the solution of KZ equations.
This model   describes the production of molecules in an atomic Fermi gas swept across an $s$-wave Feshbach resonance in the regime of a narrow resonance and sufficiently slow sweep rate $\nu$ \cite{gurarie2009}. Just as in the BCS Hamiltonian \re{bcs},  products of fermionic creation and annihilation operators $\hat c^\dagger_{j\sigma}$ and
$\hat c_{j\sigma}$, where $\sigma=\ua,\da$, are expressed in terms of Anderson pseudospins:
$2\hat s_j^z+1=\sum_\sigma\hat c^\dagger_{j\sigma}\hat c_{j\sigma}$, $\hat s_j^+=\hat c^\dagger_{j\ua}\hat c^\dagger_{j\da}$, and $\hat s_j^-=\hat c_{j\da}\hat c_{j\ua}$.
Suppose at $t\to-\infty$ the system is in the ground state. Since the bosonic energy $-\nu t\to+\infty$ in this limit, there are no  bosons.  The problem is to determine the number
of bosons $\langle \hat n_b(t)\rangle$ as $t\to+\infty$.

More generally, in this and other problems of the Landau-Zener type, we are interested in the scattering matrix that relates the state of the system at  $t=t_\mathrm{fin}$ to that at $t=t_\mathrm{in}$. For Hamiltonians linear in time, such as \eref{TC1}, $t_\mathrm{in}=-\infty$ and $t_\mathrm{fin}=+\infty$. In the case of the BCS Hamiltonian \re{bcs2} a natural choice is $t_\mathrm{in}=0^+$ and $t_\mathrm{fin}=+\infty$. The transition probability from one state at $t_\mathrm{in}$ to another at $t_\mathrm{fin}$ is modulus squared of the corresponding matrix element of the scattering matrix. For $2\times 2$ Hamiltonians linear in $t$ the problem was solved by Landau, Zener, and others back in 1932  \cite{landau,zener,Majorana,stuckelberg}, hence the name `Landau-Zener'. For time-dependent Hamiltonians with larger Hilbert spaces, even for $3\times 3$ Hermitian matrices linear in $t$, no general solution is available, but there is a  class of  models for which the multi-level version of the Landau-Zener problem is exactly solvable. 
For example, exact formulas for certain transition probabilities for the Dicke model \re{TC1} were conjectured empirically  in Ref.~\citealp{DTCM} and later justified in Ref.~\citealp{sinitsyn2017}.

The connection between solvable time-dependent models and KZ equations we establish   in this paper should be especially useful for evaluating the scattering matrix and transition probabilities.  In the theory of KZ equations, elements of the scattering matrix are known as transition functions between asymptotic solutions of these equations. There is a well-developed technique based on the general solution of KZ equations to evaluate  these transition functions  explicitly in terms of quantum $6j$-symbols \cite{varchenko}. Therefore, it should be possible to  similarly determine scattering matrices as well other quantities of interest for models we analyze in this paper. \label{6j}  We do not dwell on it further here leaving this calculation for the future.  

Well-known examples of nontrivial\footnote{`Nontrivial' in this context excludes reducible Landau-Zener models. \label{ftn1} For example, the Landau-Zener problem $g\hat J_x +\nu t \hat J_z$ for an arbitrary spin $\hat {\bf J}$ reduces to that for spin-1/2, i.e. to the original $2\times 2$ Landau-Zener problem etc., see, e.g., Ref.~\citealp{yuzbashyan-LZ} for more details.\label{ftnt1}} solvable multi-level Landau-Zener models are  Demkov-Osherov, bow-tie, and generalized bow-tie models. The first two are the following time-dependent $N\times N$ matrix Hamiltonians:
\beg
H_\mathrm{DO} =t\proj{1}+\sum_{k=2}^N \left(p_k\out{1}{k}+p_k\out{k}{1}+a_k\proj{k}\right),
\label{do}
\en
\beg
H_\mathrm{bt} = \sum_{k=2}^N \left( p_k\out{1}{k}+p_k\out{k}{1}+r_k t\proj{k} \right),
\label{bt}
\en
where, $p_k$, $a_k$, and $r_k$ are real parameters. Both describe a single level $|1\rangle$ coupled to $N-1$ other levels that are not directly coupled to each other. Diagonal matrix elements (diabatic energy levels) are linear functions of time. In the Demkov-Osherov model, also known as the equal slope model, all diabatic energies except the first one have the same slope, which has been set to zero in \eref{do} without loss of generality. In the bow-tie model the slopes are distinct and the diabatic level diagram resembles a bow-tie. 

We show in Sect.~\ref{mlz-sect} that both these models map to one of the Gaudin magnets, say $\hat H_1^G$ in \eref{gaudin}, restricted to the sector where the total spin polarization $S_z$ differs by one from its minimum, $S_z=S_z^\mathrm{min}+1=-\sum_k s_k + 1$. Note that since $\hat S_z=\sum_k \hat s_k^z$ commutes with all
$\hat H_j^G$, they break down into blocks corresponding to different eigenvalues $S_z$ of $\hat S_z$. The $N\times N$ block of $\hat H_1^G$ that corresponds to $S_z=S_z^\mathrm{min}+1$
(as well as the one for $S_z=S_z^\mathrm{max}-1$) maps to the Demkov-Osherov or bow-tie Hamiltonians via a change of variables, while similar blocks of the remaining Gaudin magnets become their commuting partners. This also allows us to construct general solutions of the non-stationary Schr\"odinger equations for these models following the same procedure as that for the KZ equations. As we will also see in Sect.~\ref{mlz-sect}, the generalized bow-tie model obtains from the bow-tie Hamiltonian \re{bt} via a simple time-independent unitary transformation, so it is not an essentially independent model and the same results as for the bow-tie model apply in this case. 

Let us also mention a generalization of the Demkov-Osherov model that describes a system of spinless fermions interacting with a time-dependent impurity level \cite{quest-LZ},
\beg
\hat H_f=t\hat n_1+\sum_{k=2}^N p_k\hat c_1^\dagger \hat c_k+p_k  c_k^\dagger \hat c_1+a_k(1-u\hat n_1)\hat n_k,\quad \hat n_j\equiv \hat c_j^\dagger \hat c_j.
\label{do-mb}
\en
Here $\hat c_j^\dagger$ and $\hat c_j$ are fermionic creation and annihilation operators and $u$ is a real parameter. It has been conjectured that the problem of determining the scattering matrix relating states at $t=\pm\infty$   is exactly solvable for this model \cite{quest-LZ}. We will show in Sect.~\ref{mb-sect} that this model too stems from Gaudin magnets \re{gaudin} and thus determine its commuting partners.

At this point it is useful to switch gears and consider instead of \esref{kz} and~\re{kzB}  an abstract set of  multi-time  Schr\"odinger equations 
\begin{equation}
\label{system1}
 i\nu \frac{\partial \Psi(\bm{x})}{\partial x_j} = \hat{H}_{j} \Psi(\bm{x}), \; \phantom{\sum} j = 0, 1, \ldots, n;
\end{equation}
 where $\bm x=(x_0, x_1,\dots,x_n)$ are
$n$ real parameters on which the Hamiltonians $\hat H_j$ depend.   Compatibility  of differential equations \re{system1} imposes severe restrictions on the choice of $\hat H_j$.
Indeed, equating mixed derivatives, $\partial_{k} \partial_{j} \Psi(\bm{x})=\partial_{j} \partial_{k} \Psi(\bm{x})$, we obtain   
\begin{eqnarray}
\label{system11}
 \partial_{j} \hat{H}_{k} - \partial_{k} \hat H_{j} - i [\hat{H}_{k}, \hat{H}_{j}] = 0, \quad k, j = 0, \ldots, n;
\end{eqnarray}
where $\partial_j\equiv \partial/\partial x_j$. These consistency conditions guarantee the existence of a  joint solution $\Psi(\bm x)$ of  \eref{system1} for any initial condition, see, e.g., Ref.~\citealp{petrat}.
Let us also introduce a nonabelian gauge field ${\cal A}(\bm{x})$,  ${\cal A}_{j} = -i\hat{H}_{j}$.  Eq.~(\ref{system11}) is then the zero curvature condition: ${\cal F}_{jk}\equiv \partial_j{\cal A}_{k}-\partial_k{\cal A}_{j}-[{\cal A}_{j},{\cal A}_{k}]=0$, meaning that the formal solution of \eref{system1} in terms of an ordered exponent
is independent of the path  connecting two fixed points  in the space of real parameters $\bm x$.

Suppose the Hamiltonians $\hat H_j$ are real symmetric. Then, the  imaginary and real parts of \eref{system11} yield 
 \beg
\label{commute-p1}
\left[\hat{H}_{j}(\bm x),\hat{H}_{k} (\bm x) \right] = 0, 
\en
\beg
\label{commute-p2}
\partial_{j} \hat{H}_{k}(\bm x) = \partial_{k} \hat{H}_{j}(\bm x),\quad j,k=0, 1,\ldots, n.
\en
These equations are useful for identifying potentially solvable time-dependent models \cite{sinitsyn2017}. For example, it is straightforward to verify that the choice $x_j=\eps_j$,  $\hat H_j=\hat H_j^G$ for $j\ge 1$,  $x_0=B$, $n=N$, and $\hat H_0=\hat H_\mathrm{BCS}$ satisfies \esref{commute-p1} and~\re{commute-p2}. The condition $[\hat H_\mathrm{BCS}, \hat H_j^G]=0$ follows from  the identities
\beg
\sum_j 2\eps_j \hat H_j^G=2B \hat H_\mathrm{BCS} - S_z^2+S_z +\sum_j {\bf s}_j^2,\quad \sum_j   \hat H_j^G=2B\hat S_z,
\label{id}
\en
where $\hat {\bf S}=\sum_j \hat {\bf s}_j$ is the total spin. As noted above, this ensures a joint solution of \esref{kz} and \re{kzB} for any initial condition. Therefore, any solution of KZ equations~\re{kz} \textit{must} also be a solution of \eref{kzB} up to a multiplicative factor  $C(B)$   independent of $\eps_j$. We will show below that  in fact $C(B)$ is also independent of $B$, i.e. is a constant, so that the solution of  \eref{kzB} coincides with the known solution of KZ equations.

In principle,  zero curvature conditions \re{commute-p1} and \re{commute-p2} seem to provide a framework for constructing  integrable time-dependent Hamiltonians  that do not necessarily  reduce to Gaudin magnets and associated KZ equations. However, so far we have not encountered  a single nontrivial example, solvable multi-level Landau-Zener problems included, for which this is the case, i.e. which is not in the Gaudin-KZ class of models. Let us emphasize that the restriction to real symmetric $\hat H_j(\bm x)$ is important here\footnote{See also the footnote on p.~\pageref{ftnt77}.}. Otherwise, there are of course numerous other realizations of
\eref{system11}, usually for $n=1$, among integrable nonlinear partial differential equations \cite{novikov-book,faddeev-book}.  Naturally,  by definition we include in the Gaudin-KZ class  other integrable versions of Gaudin magnets and corresponding KZ equations, such as trigonometric, hyperbolic, generalized to other Lie algebras instead of spin $su(2)$ etc.~\cite{gaudin,ortiz2005,babujian1994,hikami1995,santos,santos1,amico2}.

Note also that making $\eps_j$ functions
of time, $\eps_j=\eps_j(t)$, we  obtain \cite{gritsev}  the non-stationary Schr\"odinger equation for the Hamiltonian $\nu^{-1} \dot \eps_j \hat H_j^G$ directly from KZ equations~\re{kz}. 
The solution of this Schr\"odinger equation is $\Psi_\mathrm{KZ}(\bm\eps(t), B)$. What we do in this paper to derive various models listed above and solutions of their non-stationary Schr\"odinger equations from the Gaudin-KZ system is much more general. First, we map $\bm \eps$, $B$, spin magnitudes $s_j$, and associated Gaudin magnets $\hat H_j^G$ to a new set of variables $\bm x$ and associated Hamiltonians $\hat H_j$, so that the new system satisfies
the zero curvature condition~\re{commute-p2}.  In general,   the solution $\Psi(\bm x)$ of the so constructed new set of  multi-time  Schr\"odinger equations \re{system1} \textit{cannot} be obtained by performing the above mapping directly in $\Psi_\mathrm{KZ}(\bm\eps, B)$. Instead, we have to derive $\Psi(\bm x)$ anew following the same overall approach as in the case of  KZ equations\interfootnotelinepenalty=10000\footnote{Except in the case of the BCS Hamiltonian \re{bcs}, where we simply show that the existing   solution $\Psi_\mathrm{KZ}(\bm\eps, B)$ satisfies \eref{kzB}.}. That the mapping is 
zero-curvature-preserving ensures that this approach works for the new system. It is often highly nontrivial to find the proper set of variables $\bm x$ and, especially, an  appropriate construction for the solution, see, e.g., the
bow-tie example in Sect.~\ref{bt-sect}. Having done so, we make $x_j$ time-dependent to obtain the solution of  the non-stationary Schr\"odinger equation for the Hamiltonian $\nu^{-1} \dot x_j \hat H_j$ from $\Psi(\bm x)$. For example,  for the Demkov-Osherov model~\re{do},   $x_0=B-\sum_{k=2}^N s_k\eps_k^{-1}$, $p_k=\sqrt{s_k} \eps_k^{-1}$, and $x_k=a_k=-\eps_k^{-1}$ for $k\ge 2$, while $\hat H_1=H_\mathrm{DO}$. Having derived $\Psi(\bm x)$ for this case, we set $x_0=t$ to get the solution of the corresponding non-stationary Schr\"odinger equation.

\section{Time-dependent BCS Hamiltonians}
\label{bcs-sect}

Here we write down a complete set of  solutions of the evolution equation \re{kzB} for the BCS Hamiltonian \re{bcs} or, equivalently,  the  solution of the non-stationary Schr\"odinger equation for time-dependent BCS Hamiltonians \re{bcs1} and \re{bcs2}. As discussed above, this is also the solution of the KZ equations \re{kz} up to a multiplicative factor $C(B)$. Since the former is known, all we need to do is to determine $C(B)$. We will see that $C(B)$ is a $B$-independent constant and, therefore, the solution of \eref{kzB} is just the known solution of the KZ equations. Note however that various modifications of the BCS Hamiltonian \re{bcs}, e.g., adding a term $-2\mu(B) \hat S_z$ or replacing $\sum_{j,k}\hat s_j^+\hat s_k^-$ with $\hat{\bf S}^2$, do not affect the zero curvature conditions \re{commute-p1} and \re{commute-p2} while resulting in a non-constant $C(B)$.

We start with a review of the solution of KZ equations \cite{babujian2,sierra,amico1,sedrakyan,gritsev}. The  problem of determining  exact eigenvalues and eigenstates  of Gaudin magnets \re{gaudin} is  solvable. The wavefunctions are of the form
\beg
\Phi\equiv |\Phi(\bm \lambda,\bm\eps)\rangle=\prod_{\alpha=1}^M \hat L^+(\lambda_\alpha)|0\rangle,\quad \hat L^+(\lambda)=\sum_{j=1}^N \frac{\hat s_j^+}{\lambda - \eps_j},
\label{phi}
\en
where $|0\rangle$ is the minimal weight state with all spins pointing in the negative $z$-direction, $\hat s_j^z|0\rangle=-s_j|0\rangle$, $\bm\lambda=(\lambda_1,\dots,\lambda_M)$, and $\bm\eps=(\eps_1,\dots,\eps_N)$. The state $\Phi$ is a simultaneous eigenstate of $\hat H_j^G$ when $\lambda_\alpha$ satisfy a set of algebraic equations (Bethe equations). The idea is to work with the states $\Phi$  with unconstrained $\lambda_\alpha$ (off-shell Bethe Ansatz). The only additional ingredient we need is the action of $\hat H_j^G$ on these states
\beg
\hat H_j^G\Phi=h_j\Phi+ \sum_\alpha \frac{f_\alpha \hat s_j^+ \Phi_{\mathrlap{\alpha}/} }{\lam_\alpha-\eps_j},\quad \Phi_{\mathrlap{\alpha}/}=\prod_{\beta\ne\alpha} \hat L^+(\lambda_\beta),
\label{off}
\en
where $\Phi_{\mathrlap{\alpha}/}$ is the state $\Phi$ with the term $\hat L^+(\lam_\alpha)$ deleted,
\beg
h_j=-2Bs_j-\sum_{k\ne j}\frac{s_j s_k}{\eps_j-\eps_k}+\sum_\alpha\frac{s_j}{\eps_j-\lam_\alpha},
\label{h}
\en
and
\beg
f_\alpha=2B-\sum_j\frac{s_j}{\eps_j-\lam_\alpha}-\sum_{\beta\ne\alpha}\frac{1}{\lam_\beta-\lam_\alpha}.
\label{f}
\en
Note that if we were to set $f_\alpha=0$, $\Phi$ and $h_j$ would become eigenstates and eigenvalues of  $\hat H_j^G$, respectively. 

The next step is to introduce a function ${\cal S}$, known as the Yang-Yang action, defined through equations
\beg
\frac{\partial{\cal S}}{\partial\eps_j}=h_j,\quad \frac{\partial{\cal S}}{\partial\lambda_\alpha}=f_\alpha.
\label{Sdef}
\en
Explicitly we obtain
\beg
\begin{split}
{\cal S}(\bm\lam,\bm\eps)=-2B\sum_j \eps_j s_j+2B\sum_\alpha \lam_\alpha -\frac{1}{2} \sum_j\sum_{k\ne j} s_js_k \ln(\eps_j-\eps_k)+\\
\sum_j\sum_\alpha s_j\ln(\eps_j-\lam_\alpha)-
\frac{1}{2}\sum_\alpha\sum_{\beta\ne\alpha}\ln(\lam_\beta-\lam_\alpha).
\end{split}
\label{S}
\en
The solution of the KZ equations is
\beg
\Psi_\mathrm{KZ} (B, \bm\eps)= \oint_\gamma d\bm\lambda \exp\left[{-\frac{i {\cal S}(\bm\lam,\bm\eps)}{\nu}}\right]   |\Phi(\bm \lambda,\bm\eps)\rangle,\quad d\bm\lam=\prod_{\alpha=1}^M d\lam_\alpha,
\label{psi}
\en
where the closed contour $\gamma$ is such that the integrand comes back to its initial value after $\lam_\alpha$ has described it. We verify that this is indeed a solution by substituting 
$\Psi_\mathrm{KZ} (B, \bm\eps)$  into KZ equations \re{kz} and using \esref{phi}, \re{off} and \re{Sdef}. The only `trick' involved is to notice that the integral of a complete derivative with respect to any $\lambda_\alpha$ is zero due to the above property of the contour $\gamma$. Moreover, it has been shown that any solution of KZ equations \re{kz} is a linear combination of solutions of the form \re{psi} with different choices of $\gamma$, see Ref.~\citealp{varchenko} and references therein. 

To demonstrate that $\Psi_\mathrm{KZ} (B, \bm\eps)$ is also the solution of the BCS evolution equation \re{kzB} (thus also proving  $C(B)=\mbox{const}$), we apply both sides of \eref{id} to $\Psi_\mathrm{KZ}$
\beg
2i\nu\sum_j \eps_j\frac{\partial \Psi_\mathrm{KZ}}{\partial\eps_j}=2i\nu B\frac{\partial \Psi_\mathrm{KZ}}{\partial B}+A \Psi_\mathrm{KZ},
\label{C}
\en
where
\beg
A= \sum_j s_j(s_j+1)+M -\sum_j s_j-\Bigl(M-\sum_j s_j\Bigr)^2,
 \en
 and we used the fact that $\Psi_\mathrm{KZ}$ is an eigenstate of $\hat S_z$ with the eigenvalue $M -\sum_j s_j$. 
 There are several ways to prove \eref{C}. A simple way is to consider the scaling
 \beg
 \eps_j\to a\eps_j,\quad \lam_\alpha\to a\lam_\alpha,\quad B\to a^{-1}B.
 \en
 \esref{phi}, \re{S}, and \re{psi} imply
 \beg
\Phi\to a^{-M}\Phi,\quad {\cal S}\to {\cal S}+\frac{A}{2}\ln a,\quad d\bm\lam\to a^M \lam,
\en
and therefore
\beg
\Psi_\mathrm{KZ}(a^{-1}B, a\bm\eps)=\exp\left(\frac{A\ln a}{2i\nu}\right) \Psi_\mathrm{KZ}(B, \bm\eps).
\en
Differentiating this equation with respect to $a$ and then setting $a=1$, we derive \eref{C}. We have also verified \eref{C} directly by substituting \eref{psi} into \eref{C} and using \esref{off}, \re{h}, and \re{f}. Thus, $\Psi_\mathrm{KZ}(B, \bm\eps)$ is the solution of the BCS  evolution equation \re{kzB}, while $\Psi_\mathrm{KZ}(B(t), \bm\eps)$ and $\Psi_\mathrm{KZ}((\nu t)^{-1}, \bm\eps)$ solve the non-stationary Schr\"odinger equations for time-dependent BCS Hamiltonians \re{bcs1} and \re{bcs2}, respectively.

\section{Driven inhomogeneous Dicke model}
\label{D-sect}

In this section, we first describe the mapping from Gaudin magnets \re{gaudin} to the inhomogeneous Dicke model
\be
\hat{H}_\mathrm{D}= \sum_{j=1}^{N-1} \xi_j \hat{s}_j^z -\omega \hat n_b + g \!\sum_{j=1}^{N-1}(\hat{b}^{\dagger} \hat{s}^-_j + \hat{b} \hat{s}^{\dagger}_j ),
\quad \hat n_b=\hat{b}^{\dagger} \hat{b},
\label{dicke}
\ee
and then use it to derive the general solution of its non-stationary Schr\"odinger equation for $\omega=\nu t$ from the solution of KZ equations \re{psi}. 

Suppose the magnitude $s$ of one of the spins, e.g., $\hat {\bf s}_N$ diverges. We use a slightly modified Holstein-Primakoff transformation from spin to bosonic creation and annihilation operators,
\beg
\hat s^-_N=\sqrt{2s} \biggl(1-\frac{\hat n_b}{2s}\biggr)^{1/2}\hat b,\quad \hat s^+_N=\sqrt{2s}\hat b^\dagger \biggl(1-\frac{\hat n_b}{2s}\biggr)^{1/2}\!\!\!\!\!\!,\quad
\hat s_N^z=\hat  n_b - s.
\en
We only need the terms that do not vanish in the limit $s\to\infty$,
\beg 
\hat s^-_N=\sqrt{2s}\hat b+O(s^{-1/2}), \quad  \hat s^+_N=\sqrt{2s} \hat b^\dagger+O(s^{-1/2}),\quad \hat s_N^z=\hat n_b - s.
\label{HPD}
\en
Gaudin magnets \re{gaudin} now become
\beg
\begin{split}
\hat H_j^G=2B\hat s_j^z -\sum_{k\ne j}\frac{\hat {\bf s}_j\cdot \hat {\bf s}_k}{\eps_j-\eps_k}-\frac{\sqrt{2s} (\hat s_j^- \hat b^\dagger+ \hat s_j^+ \hat b ) +2(\hat n_b-s) \hat s_j^z}{2(\eps_j- \eps_N)},\\
\hat H_N^G=2B(\hat n_b-s)-\sum_j \frac{\sqrt{2s} (\hat s_j^- \hat b^\dagger+ \hat s_j^+ \hat b ) +2(\hat n_b-s) \hat s_j^z}{2(\eps_j- \eps_N)},\\
\end{split}
\label{interm}
\en
where $j=1,\dots,N-1$ and the summation over $k$ in $\hat H_j^G$  is from $k=1$ to $k=N-1$. 

Let us make the following replacements:
\beg
\eps_{j}=-\frac{\xi_j}{2g^2},\quad \eps_N=\frac{\sqrt{2s}}{2g},\quad 2B-\frac{s}{\eps_N}=\omega;\quad j\le N-1.
\label{replD}
\en
Note that $\omega=-2g^2 \eps_N+2B$, so effectively $\omega$ replaces $\eps_N$. Performing this variable change in \eref{interm}, expanding in $\sqrt{2s}$ and keeping only 
the non-vanishing terms, we obtain
\beg
\hat H_j^G\to \hat H_j^D=(\omega+\xi_j)\hat s_j^z+g(\hat s_j^- \hat b^\dagger+ \hat s_j^+ \hat b)+2g^2 \sum_{k\ne j}\frac{\hat {\bf s}_j\cdot \hat {\bf s}_k}{\xi_j-\xi_k},
\en
\beg
\hat H_N^G\to -2Bs+g\sqrt{2s} (\hat n_b+\hat S_z)-\hat H_D,\quad \hat S_z=\sum_{j=1}^{N-1} \hat s_j^z.
\en
Since the quantity $\hat n_b+\hat S_z$ commutes with $\hat H_j^D$ and the   Dicke Hamiltonian $\hat H_D$, all Hamiltonians $\hat H_j^D$ and $\hat H_D$
mutually commute. Moreover, the second zero curvature condition \re{commute-p2} holds for $x_j=\xi_j$, $\hat H_j=\hat H_j^D$ for $N-1\ge j\ge 1$, $x_0=\omega$, and $\hat H_0=\hat H_D$.

To construct the solution of the non-stationary Schr\"odinger equation for the driven inhomogeneous Dicke model $\hat H_D(t)$,
\beg
i\frac{\partial\Psi_D}{\partial t}=\hat H_D(t)\Psi_\mathrm{D}, \quad \hat{H}_\mathrm{D}(t)= \sum_{j=1}^{N-1} \xi_j \hat{s}_j^z -\nu t \hat n_b + g \!\sum_{j=1}^{N-1}(\hat{b}^{\dagger} \hat{s}^-_j + \hat{b} \hat{s}^{\dagger}_j ),
\label{dickekzD}
\en
all we need to do is to apply transformations \re{HPD} and \re{replD} to the formulas leading to the solution of the KZ equations in the previous section and take the limit $s\to\infty$.
The off-shell Bethe states \re{phi} become
\beg
\Phi\equiv |\Phi_D(\bm \lambda,\bm\xi)\rangle=\prod_{\alpha=1}^M \hat L^+(\lambda_\alpha)|0\rangle,\quad \hat L^+(\lambda)=\hat b^\dagger+g\sum_{j=1}^{N-1} \frac{\hat s_j^+}{\lambda - \xi_j},
\label{phi1}
\en
The action of $\hat H_j^D$ and $\hat H_D$ on these states is
 \beg
\hat H_j^D\Phi=h_j\Phi+ g\sum_\alpha \frac{f_\alpha \hat s_j^+ \Phi_{\mathrlap{\alpha}/} }{\lam_\alpha-\xi_j},\quad  \hat H_D\Phi=h_D\Phi+ \sum_\alpha f_\alpha \hat b^\dagger \Phi_{\mathrlap{\alpha}/}   
\label{off1}
\en
where $\Phi_{\mathrlap{\alpha}/}$ is defined as before in \eref{off} and
\beg
h_j=-\omega s_j+2g^2\sum_{k\ne j}\frac{s_j s_k}{\xi_j-\xi_k}-2g^2\sum_\alpha\frac{s_j}{\xi_j-\lam_\alpha}, \quad h_D=\sum_\alpha\lam_\alpha-\sum_k s_k\xi_k,
\label{h-D}
\en
\beg
f_\alpha=\omega-2g^2\lam_\alpha+2g^2\sum_j\frac{s_j}{\xi_j-\lam_\alpha}+2g^2\sum_{\beta\ne\alpha}\frac{1}{\lam_\beta-\lam_\alpha},
\label{f-D}
\en
where $\omega=\nu t$. We define the Yang-Yang action ${\cal S}_D$ through 
\beg
\frac{\partial{\cal S}_D}{\partial\xi_j}=h_j,\quad \frac{\partial{\cal S}_D}{\partial\omega}=h_D,\quad \frac{\partial{\cal S}_D}{\partial\lambda_\alpha}=f_\alpha,
\label{Sdef-D}
\en
which upon integration yield
\beg
\begin{split}
{\cal S}_D(\bm\lam,\bm\xi, t)= -\nu t\sum_k s_k\xi_k+\nu t\sum_\alpha\lam_\alpha +g^2 \sum_j\sum_{k\ne j} s_js_k \ln(\xi_j-\xi_k)-\\
2g^2\sum_j\sum_\alpha s_j\ln(\xi_j-\lam_\alpha)+g^2\sum_\alpha\sum_{\beta\ne\alpha}\ln(\lam_\beta-\lam_\alpha).
\end{split}
\label{S-D}
\en

Finally, the solution of the non-stationary Schr\"odinger equation \re{dickekzD} has the same form as the solution of KZ equation \re{psi}, i.e. 
\beg
\Psi_D (t, \bm\xi)= \oint_\gamma d\bm\lambda \exp\left[{-\frac{i {\cal S}_D(\bm\lam,\bm\xi, t)}{\nu}}\right]   |\Phi_D(\bm \lambda,\bm\xi)\rangle,\quad d\bm\lam=\prod_{\alpha=1}^M d\lam_\alpha.
\label{psiD}
\en
Indeed, we verify directly
\beg
\begin{split}
i\frac{\partial\Psi_D}{\partial t}-\hat H_D(t)\Psi_\mathrm{D}=- \oint_\gamma d\bm\lambda e^{\frac{-i {\cal S}_D}{\nu}}\sum_\alpha f_\alpha \hat b^\dagger \Phi_{\mathrlap{\alpha}/}
=\\
-i\nu  \sum_\alpha  \oint_\gamma \biggl(\prod_{\beta\ne\alpha} d\lam_\beta \biggr) d\lam_\alpha\frac{\partial}{\partial\lam_\alpha}\bigl( \hat b^\dagger \Phi_{\mathrlap{\alpha}/}e^{\frac{-i {\cal S}_D}{\nu}}\bigr) =0,\\
\end{split}
\en
 where we used the second equation in \re{h-D} and the last two equations in \re{Sdef-D}. By construction $\Psi_D (t, \bm\xi)$ also satisfies the remaining ($j=1,\dots, N-1$) equations~\re{system1} with $x_j=\xi_j$ and $\hat H_j=\hat H_j^D$. Let us also mention an interesting application \cite{gritsev1} of the off-shell Bethe Ansatz to the homogeneous, $\xi_j=\xi$, Dicke model with non-integrable time-dependence of the detuning $\omega$.
 
\section{Multi-level Landau-Zener models}
\label{mlz-sect}

Here we describe the mapping from a sector of Gaudin magnets \re{gaudin} to Demkov-Osherov  \cite{do}, bow-tie \cite{bow-tie}, and generalized bow-tie \cite{gbow-tie1,gbow-tie} models. 
We then derive   solutions of their non-stationary Schr\"odinger equations with the same approach as for KZ equations  in Sect.~\ref{bcs-sect}. We note that certain integral representations of these solutions have been constructed before \cite{do,be,bow-tie,gbow-tie1}. Our main point in this Section are not the solutions themselves, but to show that these models  belong to the Gaudin-KZ class and to derive their commuting partners.

The mapping proceeds in two steps. First, we note that Gaudin magnets \re{gaudin} commute with the $z$-projection of the total spin $\hat S_z$ and consider their next to minimum weight sector where $S_z=S_z^\mathrm{min}+1=-\sum_j s_j +1$. An orthonormal basis in this subspace is
\beg
|k\rangle
= \frac{\hat s_{k}^{+}|0\rangle}{\sqrt{2s_{k}}},\quad k=1,\dots,N;
\label{basic}
\en
where $|0\rangle$ is the lowest weight state $S^z=S^z_{\min}$ as in Sect.~\ref{bcs-sect}. In this basis Gaudin Hamiltonians \re{gaudin} are $N$
commuting $N\times N$ matrices
\beg
H_j^G= \tilde H_j-\biggl(2B\, s_{j}+\sum_{k\neq j}\frac{s_{j}s_{k}}{\varepsilon_{j}-\varepsilon_{k}}\biggr)\mathbb{1},\quad j=1,\dots, N;
\label{2ndterm}
\en
where $\mathbb{1}$ is the $N\times N$ identity matrix,
\beg 
\tilde H_{j}=  2B\proj{j} - \sum_{k \neq j}\frac{\g_{j}\g_{k}\out{j}{k}+\g_{j}\g_{k}\out{k}{j}-\g_{k}^{2}\proj{j}-\g_{j}^{2}\proj{k} }{\ve_{j} - \ve_{k}},  
\label{type1} 
\en
and $\gamma_k=\sqrt{s_k}$. Considering infinite-dimensional representations of the $su(2)$ algebras, where the lowest weight states are still well-defined \cite{cruz}, we treat $\gamma_k$ as arbitrary real numbers. Matrix Hamiltonians $\tilde H_j$, termed type 1 or maximal, emerge independently from a rigorous notion of quantum integrability proposed in Refs.~\cite{haile1,haile,shastry1}. Their defining feature is that this is the maximal set of   real symmetric matrices linear in a real parameter ($B$) that mutually commute, are linearly independent, and possess no $B$-independent symmetries.  Other families of mutually commuting real symmetric $N\times N$ matrices contain fewer than $N$ linearly independent members \cite{haile,shastry1}. 

It is straightforward to specialize the off-shell Bethe Ansatz equations \re{phi} -- \re{f} of Sect.~\ref{bcs-sect} to the present case.  We have
\beg
\tilde \Phi\equiv |\tilde \Phi( \lambda,\bm\eps)\rangle= \sum_{j=1}^N \frac{\gamma_j|j\rangle}{\lambda - \eps_j},
\label{phi-type1}
\en
 \beg
\tilde H_j\tilde \Phi=h_j\tilde \Phi+   \frac{f \gamma_j |j\rangle } {\lam -\eps_j},\quad h_j= \frac{\gamma_j^2}{\eps_j-\lam},\quad 
f=2B-\sum_j\frac{\gamma_j^2}{\eps_j-\lam }.
\label{off-type1}
\en
The second step is to map $\tilde H_1$ in \eref{type1} to Demkov-Osherov, bow-tie, and generalized bow-tie models.

\subsection{  Demkov-Osherov model }
\label{do-sect}

To obtain the Demkov-Osherov model, we set
\beg
\g_1=1, \quad \ve_1=0,\quad \g_k=-\frac{p_k}{a_k},\quad \ve_k=-\frac{1}{a_k},\quad 2B=t-\sum_{k=2}^N \frac{p_k^2}{a_k}, 
\label{repl-type1}
\en
where $k=2,\dots, N$. The matrix $\tilde H_1$ in \eref{type1} then turns into the Demkov-Osherov model, i.e.
\beg
\tilde H_1=H_\mathrm{DO}=t\proj{1}+\sum_{k=2}^N \left( p_k\out{1}{k}+p_k\out{k}{1}+a_k\proj{k}\right),
\label{do1}
\en
while the remaining $\tilde H_j$ are its commuting partners
\beg 
\begin{split}
\tilde H_{j} =  (t-a_j)\proj{j}-p_j\out{1}{j}-p_j\out{j}{1}  + \quad\quad\quad\quad  \\
\sum_{k \neq j}\frac{p_j p_k\out{j}{k}+ p_j p_k\out{k}{j}-p_k^{2}\proj{j}-p_j^2\proj{k} }{a_k-a_j},\\ 
\end{split}
\label{do-comm} 
\en

Importantly, in this parameterization mutually commuting matrix Hamiltonians $H_\mathrm{DO}$ and $\tilde H_j$ also satisfy the zero curvature condition \re{commute-p2}, namely,
\beg
\frac{\partial \tilde H_j}{\partial a_k}=\frac{\partial \tilde H_k}{\partial a_j},\quad \frac{\partial \tilde H_j}{\partial t}=\frac{\partial H_\mathrm{DO}}{\partial a_j},\quad j,k=2,\dots N.
\label{do0curv}
\en
We also need to redefine the spectral parameter and rescale the off-shell Bethe state in \eref{phi-type1} as follows
\beg
\eta=-\frac{1}{\lam},\quad \Phi_\mathrm{DO}=-\frac{\tilde \Phi}{\eta}.
\label{eta}
\en
Now we make replacements \re{repl-type1} and \re{eta} in \esref{phi-type1} and \re{off-type1} to obtain  
 \beg
\Phi_\mathrm{DO}\equiv|\Phi_\mathrm{DO}( \eta,\bm a)\rangle=|1\rangle- \sum_{j=2}^N \frac{p_j |j\rangle}{a_j - \eta},\quad \bm a=(a_2,\dots, a_N),
\label{phido11}
\en
\beg
\tilde H_j\Phi_\mathrm{DO}=h_j\Phi_\mathrm{DO}+  \frac{f p_j |j\rangle } {\eta -a_j},\quad  H_\mathrm{DO}\Phi_\mathrm{DO}=h_\mathrm{DO}\Phi_\mathrm{DO}+   f   |1\rangle,
\en
\beg
h_\mathrm{DO}=\eta,\quad h_j=\frac{p_j^2}{a_j}-\frac{p_j^2}{a_j-\eta},\quad f=t-\eta+\sum_{j=2}^N \frac{p_j^2}{a_j-\eta}.
\label{hdo11}
\en
As in previous sections we introduce a function ${\cal S}_\mathrm{DO}$, such that 
\beg
\frac{\partial{\cal S}_\mathrm{DO}}{\partial a_j}=h_j,\quad \frac{\partial{\cal S}_\mathrm{DO}}{\partial t}=h_\mathrm{DO},\quad \frac{\partial{\cal S}_\mathrm{DO}}{\partial\eta}=f,
\label{Sdef-DO}
\en
where $j=2,\dots,N$.  We have
\beg
{\cal S}_\mathrm{DO}(\eta,\bm a, t)=\eta t -\frac{\eta^2}{2}+\sum_{j=2}^N p_j^2\ln\left(\frac{a_j}{a_j-\eta}\right).
\en

By analogy with the solution \re{psi} of KZ equations, the solution of the non-stationary Schr\"odinger equation for the Demkov-Osherov model \re{do1} as well as the rest of equations \re{system1} for $x_j=a_j$ and $\hat H_j=\tilde H_j$ reads
\beg
\Psi_\mathrm{DO} (t, \bm a)= \oint_\gamma d\eta e^{- i {\cal S}_\mathrm{DO}(\eta,\bm a, t)}   |\Phi_\mathrm{DO}(\eta,\bm a)\rangle.
\label{psiDO}
\en
Note that replacing $\lam$ with $\eta$ and rescaling the wave-function in \eref{eta} is important for this scheme to work. These steps ensure the consistency of \esref{Sdef-DO}, $\partial h_j/\partial \eta=\partial f/\partial a_j$ and $\partial h_\mathrm{DO}/\partial \eta=\partial f/\partial t$, and that $\Psi_\mathrm{DO} (t, \bm a)$ is indeed the solution.

As discussed in the Introduction, by construction $\Psi_\mathrm{DO} (t(\tau), \bm a(\tau))$ is also the solution of the non-stationary Schr\"odinger equation 
\beg
i\frac{d\Psi}{d\tau}= H \Psi,
\en
 for the Hamiltonian  
 \beg
 H=\frac{d t}{d\tau} H_\mathrm{DO} +\sum_{j=2}^N \frac{da_j}{d\tau} \tilde H_j.
 \en
 For example, choosing linear $t(\tau)$ and $a_j(\tau)\propto\tau$, we obtain various models of the form $A+B\tau + C/\tau$, where $A$, $B$, and $C$ are $\tau$-independent $N\times N$ matrices.

\subsection{Bow-tie model}
\label{bt-sect}

The bow-tie model 
\beg
H_\mathrm{bt} = \sum_{k=2}^N \left( p_k\out{1}{k}+p_k\out{k}{1}+r_k t\proj{k} \right),
\label{bt1}
\en
obtains from the Demkov-Osherov model \re{do1} via a substitution
\beg
a_j=r_j t +t.
\en
Indeed, we find
\beg
H_\mathrm{bt} =H_\mathrm{DO} - t\mathbb{1}.
\en
This defines the mapping from one of the Gaudin magnets $\hat H_1^G$ to the bow-tie model via maximal Hamiltonians \re{type1} and the Demkov-Osherov model. Correspondingly, remaining Gaudin magnets map to commuting partners of $H_\mathrm{bt}$. Specifically, setting $a_j=r_j t+t$ in \eref{do-comm}, we get
\beg
\begin{split}
\tilde H_{j}  = -r_j t\proj{j}-p_j\out{1}{j}-p_j\out{j}{1}  +  \quad\quad\quad\quad \quad\quad \\
\sum_{k \neq j}\frac{p_j p_k\out{j}{k}+ p_j p_k\out{k}{j}-p_k^{2}\proj{j}-p_j^2\proj{k} }{(r_k-r_j)t}.
\end{split}
\label{btint} 
\en

It is not immediately clear how to identify the set of variables for which the zero curvature condition \re{commute-p2} holds.  We found the following parameterization:
\beg
r_j=\alpha_j^2,\quad p_j=\alpha_j\beta_j,\quad j=2,\dots,N.
\en
The bow-tie model in this parameterization reads
\beg
H_\mathrm{bt} = \sum_{k=2}^N \left( \alpha_k\beta_k\out{1}{k}+\alpha_k\beta_k\out{k}{1}+\alpha_k^2 t\proj{k} \right).
\label{bt2}
\en
Moreover, we have to rescale the commuting partners \re{btint},
\beg
\begin{split}
I_j =-\frac{t \tilde H_j}{\alpha_j}=\alpha_j t^2\proj{j}+\beta_jt\left(\out{1}{j}+\out{j}{1}\right)  +\quad\quad\quad\quad\quad\quad\\
 \sum_{k \neq j}\frac{\alpha_j\alpha_k\beta_j \beta_k(\out{j}{k}+   \out{k}{j})-\alpha_k^{2}\beta_k^2\proj{j}-\alpha_j^2\beta_j^2\proj{k} }{\alpha_j(\alpha_j^2-\alpha_k^2)}.\\
 \end{split}
\label{btint1} 
\en
Now the bow-tie model and this set of its commuting partners fulfill the necessary zero curvature condition, i.e.
\beg
\frac{\partial I_j}{\partial \alpha_k}=\frac{\partial I_k}{\partial \alpha_j},\quad  \frac{\partial I_j}{\partial t}=\frac{\partial H_\mathrm{bt}}{\partial \alpha_j},\quad j,k=2,\dots N.
\label{bt0curv}
\en

It is even less obvious how to make appropriate modifications in the construction of the solution of the multi-time Schr\"odinger equations of the previous subsection. Simply making the replacements $a_j=\alpha_j^2 t+t$, $p_j=\alpha_j\beta_j$ and rescaling $\tilde H_j$ in Eqs.~(\ref{phido11}-\ref{hdo11}) does not work. In addition, we have to  redefine the parameter
$\eta$ as $\eta=\kappa^2 t+t$, rescale the off-shell Bethe state $\Phi_\mathrm{DO}$, and  redefine the function $f$. We found that the following construction works:
\beg
\Phi_\mathrm{bt}\equiv|\Phi_\mathrm{bt}( \kappa,\bm \alpha)\rangle=t|1\rangle- \sum_{j=2}^N \frac{\alpha_j\beta_j |j\rangle}{\alpha_j^2 - \kappa^2},\quad \bm \alpha=(\alpha_2,\dots, \alpha_N),
\en
\beg
I_j\Phi_\mathrm{bt}=m_j\Phi_\mathrm{bt}+  \frac{f \kappa \beta_j |j\rangle } {\kappa^2 -\alpha_j^2},\quad  H_\mathrm{bt}\Phi_\mathrm{bt}=h_\mathrm{bt}\Phi_\mathrm{bt}-   f\kappa   |1\rangle,
\en
\beg
h_\mathrm{bt}=\kappa^2 t,\quad m_j=\frac{\alpha_j\beta_j^2}{\alpha_j^2-\kappa^2}-\frac{\alpha_j \beta_j^2}{\alpha_j^2+1},\quad f=\kappa t^2-\sum_{j=2}^N \frac{\alpha_j^2 \beta_j^2}{(\alpha_j^2-\kappa^2)\kappa}.
\en
As before, we define the Yang-Yang action ${\cal S}_\mathrm{bt}$ through equations 
\beg
\frac{\partial{\cal S}_\mathrm{bt}}{\partial \alpha_j}=m_j,\quad \frac{\partial{\cal S}_\mathrm{bt}}{\partial t}=h_\mathrm{bt},\quad \frac{\partial{\cal S}_\mathrm{bt}}{\partial\kappa}=f,
\label{Sdef-bt}
\en
where $j=2,\dots,N$.  These equations are consistent because $\partial m_j/\partial t=\partial h_\mathrm{bt}/\partial\alpha_j=0$, $\partial m_j/\partial \kappa=\partial f /\partial\alpha_j$,
and $\partial h_\mathrm{bt}/\partial\kappa=\partial f/\partial t$.

Integration results in
\beg
{\cal S}_\mathrm{bt}(\kappa,\bm \alpha, t)=\frac{\kappa^2 t^2}{2}-\ln\kappa\sum_{j=2}^N\beta_j^2+\frac{1}{2}\sum_{j=2}^N \beta_j^2 \ln \left( \frac{\alpha_j^2-\kappa^2}{\alpha_j^2+1}\right),
\en
and as in previous examples we write the wave-function in the form
 \beg
\Psi_\mathrm{bt} (t, \bm \alpha)= \oint_\gamma d\kappa e^{- i {\cal S}_\mathrm{bt}(\kappa,\bm \alpha, t)}   |\Phi_\mathrm{bt}(\kappa,\bm \alpha)\rangle.
\label{psibt}
\en
This wave-function solves the following system of multi-time Schr\"odinger equations
\beg
i\frac{\partial\Psi_\mathrm{bt}}{\partial t}=H_\mathrm{bt}\Psi_\mathrm{bt},\quad i\frac{\partial\Psi_\mathrm{bt}}{\partial \alpha_j}=I_j\Psi_\mathrm{bt},
\en
which includes the non-stationary Schr\"odinger equation for the bow-tie model. As before, we verify this directly. For example,
\beg
\begin{split}
 i\frac{\partial\Psi_\mathrm{bt}}{\partial \alpha_j}-I_j\Psi_\mathrm{bt}= \oint_\gamma d\kappa e^{- i {\cal S}_\mathrm{bt}}  
 \left[ i\frac{\partial |\Phi_\mathrm{bt}\rangle }{\partial\alpha_j}-\frac{f\kappa\beta_j|j\rangle}{\kappa^2-\alpha_j^2}\right]=\\
  -\oint_\gamma d\kappa e^{- i {\cal S}_\mathrm{bt}}  \left[i\frac{ (\alpha_j^2+\kappa^2)\beta_j |j\rangle}{\kappa^2-\alpha_j^2}+\frac{f\kappa\beta_j|j\rangle}{\kappa^2-\alpha_j^2}\right]=\\
  -i \oint_\gamma d\kappa \frac{\partial}{\partial\kappa} \left( e^{- i {\cal S}_\mathrm{bt}} \frac{\kappa\beta_j|j\rangle}{\kappa^2-\alpha_j^2}\right)=0\\
  \end{split}
 \en

\subsection{Generalized bow-tie model}
\label{gbt-sect}

The   generalized bow-tie model is  the following $N\times N$ matrix Hamiltonian:
\beg
\begin{array}{lcl}
\dis H_\mathrm{gbt} &=&\dis \frac{\ve}{2}\proj{a}-\frac{\ve}{2}\proj{b}+\\
\\
 &&\dis \sum_{k=3}^N \left( q_k\out{a}{k}+q_k\out{k}{a}+q_k\out{b}{k}+q_k\out{k}{b}+r_kt\proj{k}\right),\\
 \end{array}
\label{gbt}
\en
where $|a\rangle$, $|b\rangle$, and $|k\rangle$ is a set of $N$ orthonormal states. `Generalized' in the name of this model seems somewhat misleading, because it is in fact a \textit{particular case}  the usual bow-tie model after a time-independent basis change. 

Indeed, let
\beg
|1\rangle=\frac{|a\rangle+|b\rangle}{\sqrt{2}},\quad |2\rangle=\frac{|a\rangle-|b\rangle}{\sqrt{2}}.
\label{basis}
\en
The generalized bow-tie model \re{gbt} becomes
\beg
H_\mathrm{gbt} =\frac{\ve}{2} \out{1}{2}+\frac{\ve}{2} \out{2}{1}  + \sum_{k=3}^N\left( \sqrt{2}q_k\out{1}{k}+\sqrt{2}q_k\out{k}{1} +r_kt\proj{k}\right).
\label{gbt1}
\en
This is just the bow-tie Hamiltonian \re{bt1} with the following choice of parameters:
\beg
p_2=\frac{\eps}{2},\quad r_2=0,\quad p_k= \sqrt{2}q_k,\quad k=3,\dots,N.
\en
However, slight adjustments are necessary in the construction of the previous subsection, since $r_2=0$ implies $\alpha_2=0$. As a result, $I_2$  in \eref{btint1} and $\beta_2$ are
not well-defined. Instead of $I_2$ we introduce
\beg
\begin{split}
\tilde I_2=-\frac{ t \tilde H_2}{\eps}=-\frac{t}{2}(\out{1}{2}+\out{2}{1})+\frac{1}{\eps}\left(\sum_{m=3}^N  \beta_m^2\right)\proj{2}+\\
\frac{\eps}{4}\sum_{j=3}^N \frac{1}{\alpha_j^2}\proj{j}+\sum_{j=3}^N \frac{\beta_j}{ 2\alpha_j} (\out{1}{j}+\out{j}{1}).
\end{split}
\label{I2new}
\en
This is the linear in $t$ commuting partner for the generalized bow-tie model constructed in Ref.~\citealp{yuzbashyan-LZ}. In the remaining $I_j$   in \eref{btint1}
we simply set 
\beg
\alpha_2=0,\quad \alpha_2\beta_2=\frac{\eps}{2},
\en
so that
\beg
\begin{split}
I_j = \alpha_j t^2\proj{j}+\beta_jt\left(\out{1}{j}+\out{j}{1}\right)-\quad\quad\quad\quad\quad\quad \quad\\
  \frac{\eps\beta_j}{2\alpha_j^2}(\out{2}{j}+\out{j}{2})
-\frac{\beta_j^2}{\alpha_j}\proj{2}-
\frac{\eps^2}{4\alpha_j^3}\proj{j}+\quad\quad\\
 \sum_{k \neq j,2}\frac{\alpha_j\alpha_k\beta_j \beta_k(\out{j}{k}+   \out{k}{j})-\alpha_k^{2}\beta_k^2\proj{j}-\alpha_j^2\beta_j^2\proj{k} }{\alpha_j(\alpha_j^2-\alpha_k^2)},\\
 \end{split}
\label{gbtint1} 
\en
for $j=3,\dots,N$. 
Zero curvature conditions \re{bt0curv} hold as before, except ones involving $\alpha_2$ and $I_2$ are replaced with
\beg
\frac{\partial I_j}{\partial \eps}=\frac{\partial \tilde I_2}{\partial \alpha_j},\quad  \frac{\partial \tilde I_2}{\partial t}=\frac{\partial H_\mathrm{gbt}}{\partial \eps},\quad j=3,\dots N.
\label{gbt0}
\en
It is straightforward to make appropriate replacements in the remaining formulas of the previous section. Let us only give the final answer
 \beg
\Psi_\mathrm{gbt} (t, \bm \alpha, \eps)= \oint_\gamma d\kappa e^{- i {\cal S}_\mathrm{gbt}(\kappa,\bm \alpha, t)}   |\Phi_\mathrm{gbt}(\kappa,\bm \alpha, \eps)\rangle,
\label{psigbt}
\en
where $\bm\alpha=(\alpha_3,\dots,\alpha_N)$,
\beg
{\cal S}_\mathrm{gbt}(\kappa,\bm \alpha, t)=\frac{\kappa^2 t^2}{2}-\frac{\eps^2}{8\kappa^2}-\frac{\eps^2}{8}-\ln\kappa\sum_{j=3}^N\beta_j^2+\frac{1}{2}\sum_{j=3}^N \beta_j^2 \ln \left( \frac{\alpha_j^2-\kappa^2}{\alpha_j^2+1}\right),
\en
and
\beg
\Phi_\mathrm{gbt}\equiv|\Phi_\mathrm{bt}( \kappa,\bm \alpha,\eps)\rangle=t|1\rangle+\frac{\eps}{\kappa^2}|2\rangle- \sum_{j=3}^N \frac{\alpha_j\beta_j |j\rangle}{\alpha_j^2 - \kappa^2}.
\en

\section{Many-body extension of the Demkov-Osherov model}
\label{mb-sect}

In this section we analyze  a many-body extension of the Demkov-Osherov model to a system of spinless fermions interacting with a time-dependent impurity level \cite{quest-LZ}.
The Hamiltonian is
\beg
\hat H_f=t\hat n_1+\sum_{k=2}^N p_k\hat c_1^\dagger \hat c_k+p_k  c_k^\dagger \hat c_1+a_k(1-u\hat n_1)\hat n_k,\quad \hat n_j\equiv \hat c_j^\dagger \hat c_j.
\label{do-mb11}
\en
 When the interaction $u=0$ and there is only one fermion in the system, this is the Demkov-Osherov Hamiltonian \re{do1}. We will derive this model together with its commuting
 partners from Gaudin magnets. 
 We, however, will not pursue a solution of its non-stationary Schr\"odinger equation here as we did for the other models.
 
 We start by casting Gaudin magnets \re{gaudin} into a more convenient form,
 \beg
\hat H_j^G=2B\hat s_j^z-\sum_{k\ne j}\frac{\frac{1}{2}(\hat s_j^+\hat s_k^- +\hat s_j^- \hat s_k^+) + \hat s_j^z \hat s_k^z}{\eps_j-\eps_k},\quad [\hat H_i^G, \hat H_j^G]=0.
\label{gaudin11}
\en
Let us represent spins in terms of bosons via a variant of Holstein-Primakoff transformation,
\beg
\hat s^-_j=\sqrt{2s_j} \biggl(1-\frac{\hat b_j^\dagger \hat b_j}{2s_j}\biggr)^{1/2}\hat b_j,\quad \hat s^+_j=\sqrt{2s_j}\hat b_j^\dagger \biggl(1-\frac{\hat b_j^\dagger\hat b_j}{2s_j}\biggr)^{1/2} \!\!\!\!\!\!,\quad
\hat s_j^z=\hat  b_j^\dagger \hat b_j - s_j.
\en
Expansion in inverse spin magnitudes yields,
\beg
\hat s_j^+\hat s_k^- =2\sqrt{s_j s_k} \left( \hat b_j^\dagger \hat b_k -\frac{ (\hat b_j^\dagger)^2 \hat b_j \hat b_k}{4 s_j} -\frac{ \hat b_j^\dagger\hat b_k^\dagger (\hat b_k)^2}{4 s_j}+\dots\right).
\label{expss}
\en
Since all terms in the expansion of $ \hat H_j^G$ contain even number of bosonic creation and annihilation operators, replacing bosons with fermions does not affect the commutation relation $[\hat H_i^G, \hat H_j^G]=0$. Similarly, we are free to replace $\sqrt{s_j}$ with arbitrary real numbers $\gamma_j$, i.e.
\beg
 \hat b_j^\dagger\to \hat c_j^\dagger,\quad \hat b_j\to \hat c_j,\quad \sqrt{s_j}=\gamma_j.
 \en 
Then, all terms in brackets in \eref{expss} except the first one vanish and we obtain
\beg
\hat H_j^G=\hat H_j^{(0)}+\hat H_j^{(1)}+\hat H_j^{(2)},
\en
where $\hat H_j^{(0)}$ are constants given by the second term on the right hand side of \eref{2ndterm} and
\beg
\hat H_j^{(1)}=2B\hat n_j - \sum_{k \neq j}\frac{\g_{j}\g_{k}(\hat c_j^\dagger\hat c_k+  c_k^\dagger\hat c_j) -\g_{k}^{2}\hat n_j  -\g_{j}^{2}\hat n_k}{\ve_{j} - \ve_{k}},
\en
\beg
 \hat H_j^{(2)}=- \sum_{k \neq j}\frac{\hat n_j \hat n_k }{\ve_{j} - \ve_{k}}.
 \label{hj2}
\en
We have
\beg
[\hat H_i^{(1)}, \hat H_j^{(1)}]=0,\quad [\hat H_i^{(1)}+\hat H_i^{(2)}, \hat H_j^{(1)}+\hat H_j^{(2)}]=0.
\label{orders}
\en
The first relation holds because it corresponds to the leading order in the expansion in inverse spin magnitudes. Independently, it follows from the fact that $\hat H_j^{(1)}$ are the type 1 Hamiltonians \re{type1} dressed with fermions, i.e. $\hat H_j^{(1)}=\sum_{l,m} (\tilde H_j)_{lm} \hat c_l^\dagger\hat c_m$. Further, \esref{orders} imply that operators $\hat K_j=H_j^{(1)}+u\hat H_j^{(2)}$ mutually commute. Note that we equivalently acquire a parameter $u$ in front of $\hat H_j^{(2)}$ via a simple rescaling $\eps_i\to u^{-1}\eps_i$ and $\gamma_i\to u^{-1/2}\gamma_i$. 

Thus, we have derived the
following set of mutually commuting Hamiltonians from Gaudin magnets:
\beg
 \hat K_j=2B\hat n_j - \sum_{k \neq j}\frac{\g_{j}\g_{k}(\hat c_j^\dagger\hat c_k+  c_k^\dagger\hat c_j) -\g_{k}^{2}\hat n_j  -\g_{j}^{2}\hat n_k+u\hat n_j \hat n_k }{\ve_{j} - \ve_{k}}.
\en
Finally, under the transformation \re{repl-type1}, $\hat K_1$ turns into the fermion model \re{do-mb11}, while the remaining $\hat K_j$ are its commuting partners. We read them off  \eref{do-comm}, replacing $|i\rangle\to \hat c_i$ and adding the term \re{hj2} proportional to $u$, where $(\eps_j-\eps_k)^{-1}=a_j a_k (a_k-a_j)^{-1}$. We obtain
\beg 
\begin{split}
\tilde H_{j} =  (t-a_j)\hat n_j-p_j(\hat c_1^\dagger \hat c_j-\hat c_j^\dagger \hat c_1)  + \quad\quad\quad\quad\quad\quad\quad\quad   \\
\sum_{k \neq j}\frac{p_j p_k(\hat c_j^\dagger\hat c_k+  c_k^\dagger\hat c_j)-p_k^{2}\hat n_j-p_j^2\hat n_k -u a_j a_k \hat n_j \hat n_k}{a_k-a_j}.\\ 
\end{split}
\label{do-comm11} 
\en

\section{Discussion}

We have constructed complete sets of solutions of the non-stationary Schr\"odinger equation for a number of time-dependent models. Among them are two interacting 
many-body Hamiltonians -- the driven inhomogeneous Dicke \re{TC1} and the time-dependent BCS \re{bcs1} models.   The former describes molecular production in an atomic Fermi gas swept through a narrow Feshbach resonance. The latter include, in particular, the BCS Hamiltonian \re{bcs2} with a coupling constant inversely proportional to time as well as periodically driven (Floquet) BCS models, e.g., for $B=B_0 \cos \nu t$ in \eref{bcs1}.

It is instructive to assess our results in the context of the theory of exactly solvable multi-level Landau-Zener problems. These are Hamiltonians
of the form $A+B t$, where $A$ and $B$ are time-independent real-symmetric matrices of arbitrary size. The problem is considered solvable if one is able to determine  transition probabilities between states at $t=\pm\infty$ explicitly in terms of matrix elements of $A$ and $B$.  Over the years, only a few nontrivial\footnote{See the footnote on p.~\pageref{ftn1}} solvable Hamiltonians have been identified. The  main ones are the Demkov-Osherov \re{do}, bow-tie \re{bt}, and the generalized bow-tie \re{gbt} models. It
turns out that the key special property of these models is the presence of nontrivial commuting partners~\cite{yuzbashyan-LZ}. Moreover,  we have demonstrated in this paper that all these models map to a particular sector of Gaudin magnets~\re{gaudin} and presented a new, improved construction of their commuting partners.
 
Building on the presence of commuting partners in  nontrivial solvable Landau-Zener models, Ref.~\citealp{sinitsyn2017}  proposed a method of determining the transition probabilities based on zero curvature conditions \re{commute-p1} and \re{commute-p2}. Here we have seen that the Demkov-Osherov, bow-tie, and the generalized bow-tie models as well as the inhomogeneous Dicke Hamiltonian indeed satisfy these conditions. Moreover,   we solved  the non-stationary Schr\"odinger equation for these models by paralleling the off-shell Bethe Ansatz  solution of the Knizhnik-Zamolodchikov equations. The Demkov-Osherov and bow-tie models give rise via the procedure outlined at the end of Sect.~\ref{do-sect} (see also Ref.~\citealp{sinitsyn2017})
to derivative solvable Hamiltonians of the form $A+Bt$ and $A+Bt +C/t$, which look very similar to those introduced in Refs.~\cite{LZC,LZC1,LZC2,chen-largeLZ}.
However, most interesting would be to identify nontrivial solvable Hamiltonians of this form\footnote{Here we mean `scalable' models \label{ftnt77} uniformly defined for arbitrary matrix size or particle (spin) number, rather than models integrable only for a given fixed matrix size or particle (spin) number.} 
that do not reduce to Gaudin magnets and whose non-stationary Schr\"odinger equations are not of Knizhnik-Zamolodchikov type.

More work is needed to extract various physical information from exact solutions of the non-stationary Schr\"odinger equations presented in this paper. For example,
an interesting quantity to evaluate in the Dicke model   is the quantum mechanical average number of bosons $\langle \hat n_b(t)\rangle$ as a function of time. Several  predictions are available for this quantity in the limit $t\to+\infty$ for both homogeneous \cite{altland2008,altland2009,itin2009,itin2010,sun2016} and inhomogeneous \cite{gurarie2009} Dicke models, such as, e.g.,  the breakdown of the adiabaticity \cite{altland2009}. A potentially useful tool for such a calculation are the matrix elements of $\hat n_b$ and spin operators between the off-shell Bethe Ansatz states for the inhomogeneous Dicke and Gaudin Hamiltonians \cite{faribault2012,faribault2014}. It might be then possible to obtain simple expressions for large time asymptotes of $\langle \hat n_b(t)\rangle$
and other quantities of interest in
the thermodynamic limit.  Let us also note in this connection the semiclassical asymptotic expansion for the solution of Knizhnik-Zamolodchikov 
equations \cite{varchenko}.   Another application is to determine the scattering matrix and related observables with the help of the machinery developed for evaluating transition functions between asymptotic solutions of Knizhnik-Zamolodchikov 
equations as discussed on p.~\pageref{6j} of the Introduction.

\section*{Acknowledgment}

This work was  supported  by the National Science Foundation Grant DMR-1609829.

\end{document}